\documentclass[%
 aip,
cp,  
 amsmath,amssymb,
 reprint,%
]{revtex4-2}

\usepackage{graphicx}
\usepackage{dcolumn}
\usepackage{bm}

\usepackage[utf8]{inputenc}
\usepackage[T1]{fontenc}
\usepackage{mathptmx} 
\usepackage{color,amsmath}%
\usepackage{ccaption} 

\newcommand*\Laplace{\mathop{}\!\mathbin\bigtriangleup}

\begin{document}

\title{Origin of anisotropic diffusion in Turing Patterns}

\author{H. Koibuchi} 
 \email[Corresponding author: ]{koibuchi@gm.ibaraki-ct.ac.jp; koibuchi@ibaraki-ct.ac.jp}
\author{F. Kato}%
\affiliation{
  National Institute of Technology (KOSEN), Ibaraki College, Hitachinaka, Japan.
}
\author{G. Diguet}
\affiliation{%
Micro System Integration Center, Tohoku University, Sendai, Japan
}%
\author{T. Uchimoto}
\affiliation{%
Institute of Fluid Science (IFS), Tohoku University, Sendai, Japan
}%
\affiliation{ELyTMaX, CNRS-Universite de Lyon-Tohoku University, Sendai, Japan}

\date{\today} 

\begin{abstract}
In this paper, we numerically study Turing patterns by the Finsler geometry (FG) modeling technique on thermally fluctuating triangular lattices, which are often used for modeling cell membranes or lipid membranes,  focusing on the origin of diffusion anisotropy. The FG modeling prescription allows us to assume direction-dependent diffusion described by Laplacian. To implement such diffusion anisotropy in the FG modeling, we need an internal degree of freedom (IDF), which depends on direction and position and  is controlled by some external forces or stimuli. For such a direction-dependent IDF, we use velocity directions corresponding to thermal fluctuations of the lattice vertices.  We find that anisotropic Turing patterns emerge in the direction along which vertices fluctuate. In the simulations, direction-dependent diffusion coefficients are unnecessary for input, and instead, the IDF aligns  the direction of vertex fluctuation  along a direction implemented by external stimuli. Our results and techniques provide insight into the origin of diffusion anisotropy connected to Turing patterns.
\end{abstract}

\maketitle

\section{\label{sec:intro}Introduction} 
\begin{figure}[h!]
\centering{}\includegraphics[width=14.5cm]{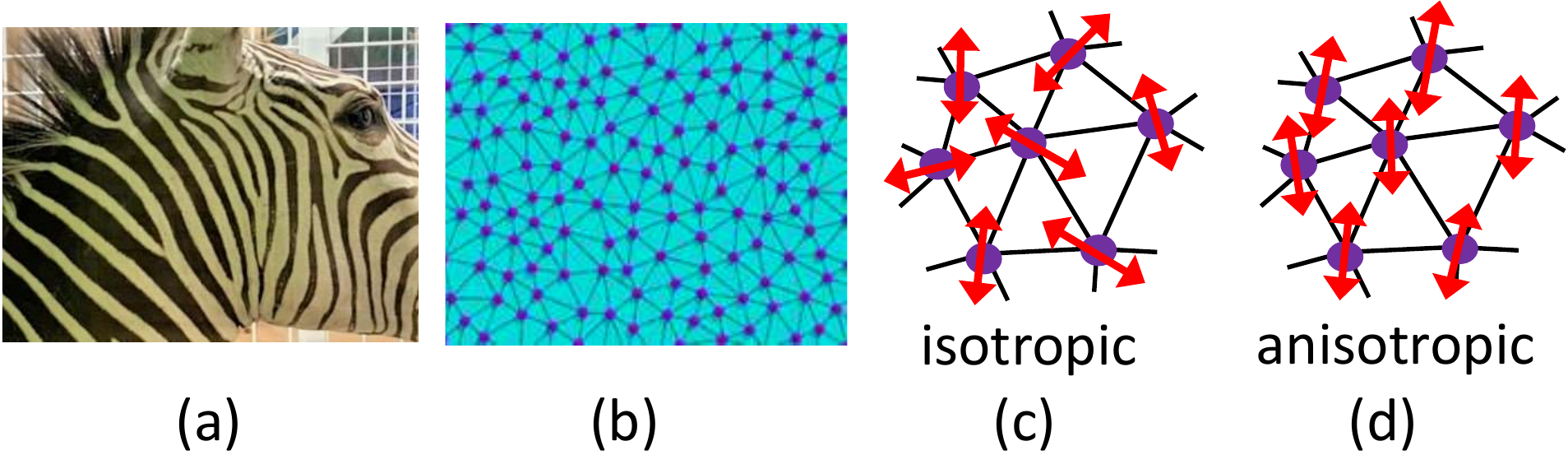}
\caption{(a) An example of a Turing pattern, (b) two-dimensional triangulated lattice, where small balls represent lipid molecules or lumps of lipids, (c) omni-directional or isotropic fluctuations of lipids corresponding to isotropic diffusion, and (d) aligned or anisotropic fluctuations of lipids corresponding to anisotropic diffusion, where the arrows represent the fluctuation directions. The fluctuation is in-plane directions only. \label{fig-1}  }
\end{figure}

Turing patterns  (Fig. \ref{fig-1}(a)) are described by a set of partial differential equations composed of diffusion and reaction terms \cite{Turing-TRSL1952}. This equation has an unstable plane wave solution in an unstable environment called Turing instability \cite{Kuramoto-Seibutsu1979,Chiu-PRE2008,Kondo-LN1995,Dziekan-JCP2012}. In this situation, the appearance of isotropic Turing patterns can be checked numerically with suitable pairs of diffusion constants. Moreover, if the diffusion constants are direction-dependent, anisotropic patterns such as those on zebra emerge.  However, the origin of this anisotropy in diffusion coefficients is unclear. 

We numerically study this problem  on two-dimensional triangular lattices of fixed connectivity  \cite{KANTOR-NELSON-PRA1987,Gompper-Kroll-PRA1992,KOIB-PRE-2005} and dynamically-triangulated fluid lattices  \cite{Ho-Baum-EPL1990}  (Fig. \ref{fig-1}(b)). Triangulated surfaces have been used to study the shape morphology of lipid membranes or cell membranes \cite{Peliti-Leibler-PRL1985,NELSON-SMMS2004} by assuming a Helfrich Hamiltonian, which is the curvature energy of membranes in three-dimensional space \cite{HELFRICH-1973}. Triangular meshes are an extension of the linear chain model for polymers \cite{Doi-Edwards-1986}. In such a model for membranes, triangle vertices correspond to lipid molecules. For this reason,  fluctuating triangular lattices are suitable for studying the origin of zebra patterns. 
 We use the Finsler geometry (FG) modeling technique, which connects anisotropy in vertex diffusion  with anisotropy in Laplacian. To connect these two different anisotropies, we need to include internal degrees of freedom (IDF) for distance anisotropy on the triangulated lattices in the FG modeling. 
Another possible origin of IDF is bacterial swarming or navigation on surfaces \cite{Ariel-etal-NJP2013,Ariel-etal-NatCom2015}. The bacterial cell size is $0.4 (\rm{\mu m})$; therefore, the cells can thermally fluctuate. However, this movement of bacteria is mainly activated by some environmental conditions and is not always described by the standard diffusion equation \cite{Ariel-etal-NJP2013,Ariel-etal-NatCom2015}. Therefore, we study in this paper whether anisotropic Turing patterns are described by diffusion anisotropy, which is controlled by some IDF connected to the diffusion itself.
 
 The problem is to find the content or identity of IDF. One possible answer is that this IDF corresponds to the direction of thermal fluctuations of vertices of the lattice.  The movement of vertices is characterized by velocity, which has a directional degree of freedom and plays a role in IDF  (Fig. \ref{fig-1}(c),(d)). Without a specific condition, thermal fluctuations, including the directions, are at random leading to isotropic diffusion.  Therefore, anisotropic Turing patterns are expected to emerge only for aligned velocity along a direction oriented by some external stimuli or forces. This velocity alignment, representing an actual anisotropic diffusion of lattice vertices, causes anisotropic Turing patterns on both fixed connectivity and fluid surfaces. 

\section{Models}
\subsection{FitzHugh-Nagumo equation and anisotropic Turing patterns}
Let $\phi(x,y)$ and $\psi(x,y)$ be scalar functions at two-dimensional space point $(x,y)$ satisfying a set of partial differential equations called the FitzHugh-Nagumo equation:
\begin{eqnarray}
\begin{split}
\label{FN-eq-Eucl}
\frac {\partial \phi}{\partial t}=D_\phi {\Laplace}\phi +f(\phi ,\psi), \quad \frac {\partial \psi}{\partial t}=D_\psi  \Laplace \psi+\gamma g(\phi,\psi), \quad f=\phi -\phi ^3-\psi, \;
\quad g= \phi -\alpha \psi -\beta,
\end{split}
\end{eqnarray}
where Laplace operator is defined by $\Laplace=\frac{\partial^2 }{\partial x^2}+\frac{\partial^2}{\partial y^2}$, and $D_\phi$ and $D_\psi$ are the diffusion constants. Other parameters $f$ and $g$ are called reaction terms, and $\alpha$, $\beta$ and $\gamma$ are constants. 

\begin{figure}[h!]
\centering{}\includegraphics[width=14.5cm]{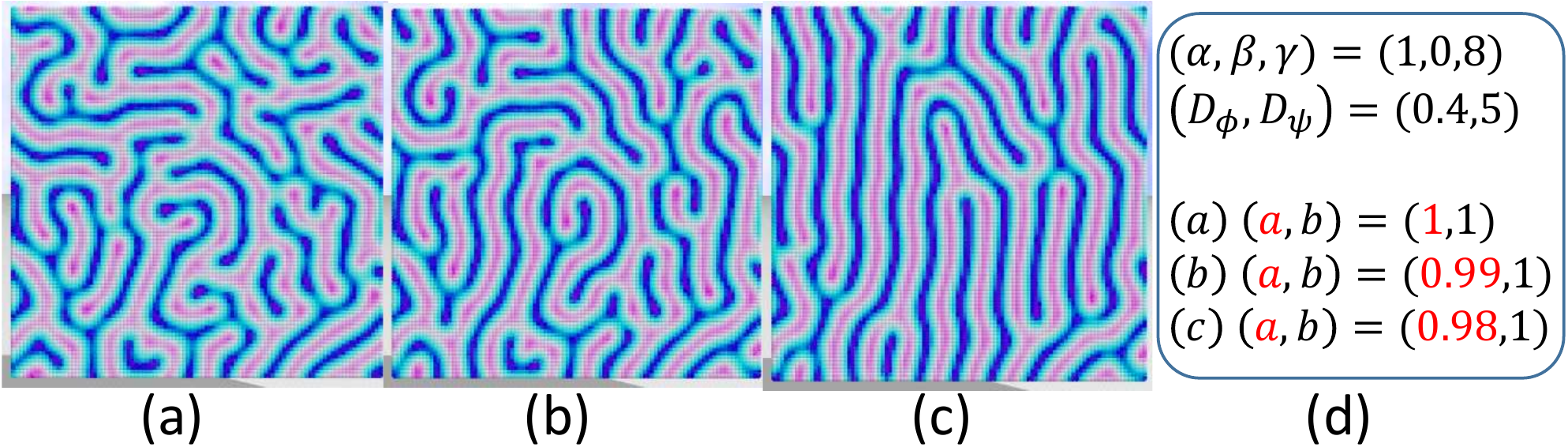}
\caption { (a) An isotropic configuration obtained under the condition $(a,b)\!=\!(1,1)$ in the modified Laplacian in Eq. (\ref{modified-Laplacian}), and anisotropic configurations obtained at (b) $(a,b)\!=\!(0.99,1)$, (c) $(a,b)\!=\!(0.98,1)$ with (d) the other parameters assumed in the FitzHugh-Nagumo equation in Eq. (\ref{FN-eq-Eucl}). The size of the regular square lattice is $N\!=\!100^2$.
\label{fig-2} }
\end{figure}
It is well-known that a time-dependent plane wave solution is close to a constant solution, and  the time-dependent solution possesses an unstable behavior corresponding to Turing patterns  in a specific range of constants. We are interested in anisotropic Turing patterns, and therefore, we modify the Laplacian to 
\begin{eqnarray}
\label{modified-Laplacian}
\Laplace \phi \to a\frac{\partial^2 \phi }{\partial x^2}+(2-a)\frac{\partial^2\phi}{\partial y^2},\; (0<a<2), \quad
\Laplace \psi \to b\frac{\partial^2 \psi }{\partial x^2}+(2-b)\frac{\partial^2\psi}{\partial y^2},\; (0<b<2) 
\end{eqnarray}
both of which restore to the isotropic $\Laplace$ in the limit of $a\!\to\!1$, $b\!\to\! 1$. Here, we use the regular square lattice of size $N\!=\!100^2$ to solve the steady-state solutions satisfying the conditions ${\partial \phi }/{\partial t}\!=\!0$ and ${\partial \psi }/{\partial t}\!=\!0$ in Eq. (\ref{FN-eq-Eucl}).  
Figures \ref{fig-1}(a)--(c) show (a) isotropic and (b),(c) anisotropic patterns, where the assumed parameters are shown in Fig. \ref{fig-1}(d). We find that the anisotropic pattern in Fig. \ref{fig-1}(c) is more evident than that in Fig. \ref{fig-1}(b). The reason for the appearance of such an anisotropic pattern is that $D_\phi$ is effectively direction-dependent such that $D_\phi\to(D_\phi^x,D_\phi^y)=D_\phi (a, 2\!-a\!)\!=\!(0.98D_\phi, 1.02D_\phi )$ for $a\!=\!0.98$. Note that $(a, 2\!-a\!)$ or  $(b, 2\!-b\!)$ is the reason for the  anisotropy in diffusion constants $(D_\phi^x,D_\phi^y)\!=\!(aD_\phi,(2-a)D_\phi)$ or $(D_\psi^x,D_\psi^y)\!=\!(bD_\psi,(2-b)D_\psi)$. As  a result,  $(D_\phi^x,D_\phi^y)$ changes from isotropic to anisotropic such that $(D_\phi^x,D_\phi^y)\!=\!(D_\phi,D_\phi)\to (D_\phi^x,D_\phi^y)\!=\!(0.98D_\phi, 1.02D_\phi)$ if $(a,b)$ change as $(a,b)\!=\!(1,1)\!\to\!(a,b)\!=\!(0.98,1)$. 
Here, we assume anisotropy only in the variable $\phi$. However, it is still unclear the origin or mechanism for $a$ to deviate from $a\!=\!1$ to  $a\!=\!0.98$.

\subsection{Finsler geometry modeling of anisotropic diffusion}
To find the reason for $(D_\phi^x,D_\phi^y)$ being anisotropic, we use the FG modeling technique. In this FG modeling,  a new IDF $\vec{\tau}$ is introduced for the variables $\phi$ and $\psi$ in Eq. (\ref{FN-eq-Eucl}), and the diffusion constant $D_\psi$ also becomes isotropic for the variable $\psi$. The  variable $\vec{\tau}$ treated as a dynamical variable in addition to $\phi$ and $\psi$, is updated by Monte Carlo (MC) simulations. For this purpose, we need a Hamiltonian depending on $\vec{\tau}$.

Thus, we start with the following Hamiltonian:
\begin{eqnarray}
\label{cont-Hamiltonian}
S=S_\phi+S_\psi,\quad S_{\phi}= \frac{1}{2}\int \sqrt{g}d^2x \,g^{ab}\frac{\partial \phi }{\partial x^a}\frac{\partial \phi }{\partial x^b}, \quad S_{\psi}= \frac{1}{2} \int \sqrt{g}d^2x \,g^{ab}\frac{\partial \psi }{\partial x^a}\frac{\partial \psi }{\partial x^b}, 
\end{eqnarray}
The $S_{\phi}$and  $S_{\psi}$ in $S$ are written by using the inverse $g^{ab}$ and the determinant $g$ of the Finsler  metric $g_{ab}$, where $(x^1,x^2)$ represents a local coordinate of the surface. 
 By the variation principle, we have a differential equation for $\phi$ and $\psi$ associated with the operator expression 
\begin{eqnarray}
\label{FN-eq}
 \Laplace =\frac{1}{\sqrt{g}}\frac{\partial}{\partial x_a}\left(\sqrt{g}g^{ab}\frac{\partial  }{\partial x^b}\right),
\end{eqnarray}
which is a Laplacian  on curved surfaces. In this expression, $g$ and $g^{ab}$ depend on $\vec{\tau}$, which is also dependent on the local coordinate directions $x^1$ and $x^2$.  
Therefore,  the operator $\Laplace$ in Eq. (\ref{FN-eq}) can be written as  $\Laplace_\phi\!=\! a_x(\tau)\frac{\partial^2 }{\partial x^2}\!+\! a_y(\tau)\frac{\partial^2 }{\partial y^2} \!+\! \cdots$ and  $\Laplace_\psi\!=\! b_x(\tau)\frac{\partial^2 }{\partial x^2}\!+\! b_y(\tau)\frac{\partial^2 }{\partial y^2} \!+\! \cdots$, where  $a_x(\tau)$, $a_y(\tau)$ and $b_x(\tau)$, $b_y(\tau)$ play roles in $a$, $2\!-\!a$ and $b$, $2\!-\!b$ in $\Laplace_\phi$ and $\Laplace_\psi$ in Eq. (\ref{modified-Laplacian}), respectively. The detailed information on the discretization will be reported elsewhere.  

\begin{figure}[h!]
\centering{}\includegraphics[width=12.5cm]{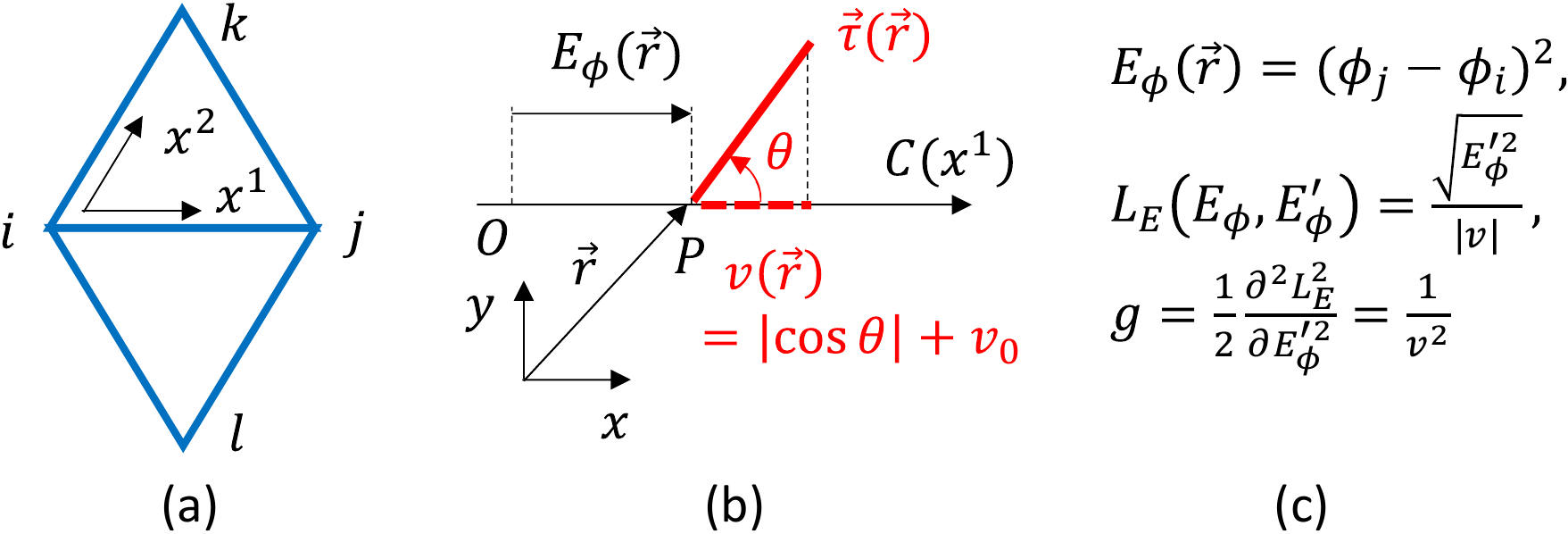}
\caption {(a) Two triangles $ijk$ and $ilj$ sharing bond $ij$,  and a local coordinate $(x^1,x^2)$ of triangle $ijk$  at vertex $i$,  (b) a line $C(x^1)$ along the $x^1$ axis, one-dimensional energy $E_\phi(\vec{r})$ for Finsler length along $C(x^1)$ and the corresponding IDF $\vec{\tau}$ at $\vec{r}$, (c) Expressions of $E_\phi(\vec{r})$, Finsler function $L_E$ for $\phi$ and the corresponding Finsler metric $g$ along $C(x^1)$. The definition of  $v_{ij}$ for $\phi$ is different from that of $v_{ij}$ for $\psi$ as shown in Eq. (\ref{one-dim-Hamiltonian}).
\label{fig-3} }
\end{figure}
Using $S_{\phi}$ and $S_{\psi}$, we introduce two different Finsler time lengths in the local coordinate system on the lattice. First we introduce one-dimensional discrete energies $E_\phi$ and $E_\psi$ corresponding to $S_{\phi}$ and $S_{\psi}$ to define the Finsler length $v_{ij}$ for $\phi$ and $\psi$
\begin{eqnarray}
\label{one-dim-Hamiltonian}
\begin{split}
&E_{\phi}= \sum_{ij\in C(x^1)}\left(\phi_j-\phi_i\right)^2, \quad  v_{ij}^\phi=|\vec{e}_{ij}\cdot \vec{\tau}_i|+v_0\\
& E_{\psi}= \sum_{ij\in C(x^1)}\left(\psi_j-\psi_i\right)^2, \quad  v_{ij}^\psi=\sqrt{1-(\vec{e}_{ij}\cdot \vec{\tau}_i)^2}+v_0
\end{split}
\end{eqnarray}
where $ij$ denotes a bond along the line $C(x^1)$ representing a local coordinates axis (Figs. \ref{fig-3}(a),(b)). This energy is used to define the Finsler time length $s$ along the $x^1$ axis by using the Finsler function  $L_E\!=\!\sqrt{E^{\prime 2}}/|v|$ such that $L_E\!=\!ds/dx^1$, where $v$ is a velocity or Finsler length along the $x^1$ axis (Fig. \ref{fig-3}(c))  (see also Ref.\cite{SElHog-etal-RIP2022} for more detailed information on this part).  To define $v$, we introduce a variable $\vec{\tau} (\in S^1/2)$, which will be described in the following subsection. The Finsler length along the $x^2$ axis is defined in the same way as that along the $x^1$ axis and is implemented in the model through the metric tensor $g_{ab}$ such that $(g_{11},g_{22})\!=\!(v_{12}^{-2},v_{13}^{-2})$.  Finsler lengths $v_{ij}$ for $\phi$ and  $\psi$ denoted by  $v_{ij}^\phi$ and $v_{ij}^\psi$ are defined by using the unit tangential vector $\vec{e}_{ij}$ from $i$ to $j$ and $\vec{\tau}_i$ at $i$ as in Eq. (\ref{one-dim-Hamiltonian}). Note that $v_{ij}^\phi$ ($v_{ij}^\psi$) is long along the direction parallel (perpendicular) to the $\vec{e}_{ij}$ direction and is controlled by some external force $\vec{B}$ because the direction of $\vec{\tau}_i$ is controlled by $\vec{B}$, which will be introduced below.

Using those obtained metric tensors corresponding to IDF $\vec{\tau}$  including several additional energies defined on triangular lattice, we obtain the discrete Hamiltonian  \cite{ICMsquare2020-JPCconf2022} 
\begin{eqnarray}
\label{discrete-Hamiltonian}
\begin{split}
&S(\vec{r};\vec{\tau})=S_1+D_\phi S_{\phi}+D_\psi S_{\psi}+\lambda_\phi S_0^\phi+\lambda_\psi S_0^\psi+S_B^\phi, \\
&S_1=\sum_{ij}\ell_{ij}^2,\quad \ell_{ij}^2\!=\!(\vec{r}_i\!-\!\vec{r}_i)^2,\\
&S_{\phi}=\sum_{i}\sum_{j(i)}\gamma_{ij}^\phi\left(\phi _j-\phi _i\right)^2, \quad S_{\psi}=\sum_{i}\sum_{j(i)}\gamma_{ij}^\psi\left(\psi _j-\psi _i\right)^2, \quad S_B^\phi=-\sum_i\left(\vec{\tau}_i\cdot {\vec B}\right)^2, \\
&\gamma_{ij}^\phi=\frac{1}{6}\left(\frac{v_{ij}^\phi}{v_{ik}^\phi}+\frac{v_{ji}^\phi}{v_{jk}^\phi}+\frac{v_{ij}^\phi}{v_{il}^\phi}+\frac{v_{jl}^\phi}{v_{jl}^\phi}\right), \quad
\gamma_{ij}^\psi=\frac{1}{6}\left(\frac{v_{ij}^\psi}{v_{ik}^\psi}+\frac{v_{ji}^\psi}{v_{jk}^\psi}+\frac{v_{ij}^\psi}{v_{il}^\psi}+\frac{v_{jl}^\psi}{v_{jl}^\psi}\right),\\
\end{split}
\end{eqnarray}
where  $ij$ in $\sum_{ij}$ in $S_1$ denotes the bond connecting two neighboring vertices $i$ and $j$ and $\ell_{ij}^2\!=\!(\vec{r}_i\!-\!\vec{r}_i)^2$ is the bond length squares. $\sum_{i}\sum_{j(i)}$ in $S_{\phi}$ and $S_{\psi}$ denotes the sum over vertex $i$ and the sum over bonds $j(i)$ connected to $i$.  
The first term $S_1$ is the bond potential, which makes the mean value of $\ell_{ij}$ uniform.   The symbol ${\vec B}$  in the term $S_B^\phi$  denotes  some external stimuli or forces aligning the IDF $\vec{\tau}_i$. We should note that $\vec{\tau}_i$ is forced to align along the direction of ${\vec B}$ when $|{\vec B}|\!\not=\!0$, and as a result, thermal fluctuations of $\vec{r}$ also align along ${\vec B}$. More detailed information on the Hamiltonian, including the discretization of $S_{\phi}$ and $S_{\psi}$, will be reported elsewhere. 

\subsection{Hybrid numerical technique}
\begin{figure}[h!]
\centering{}\includegraphics[width=11.5cm]{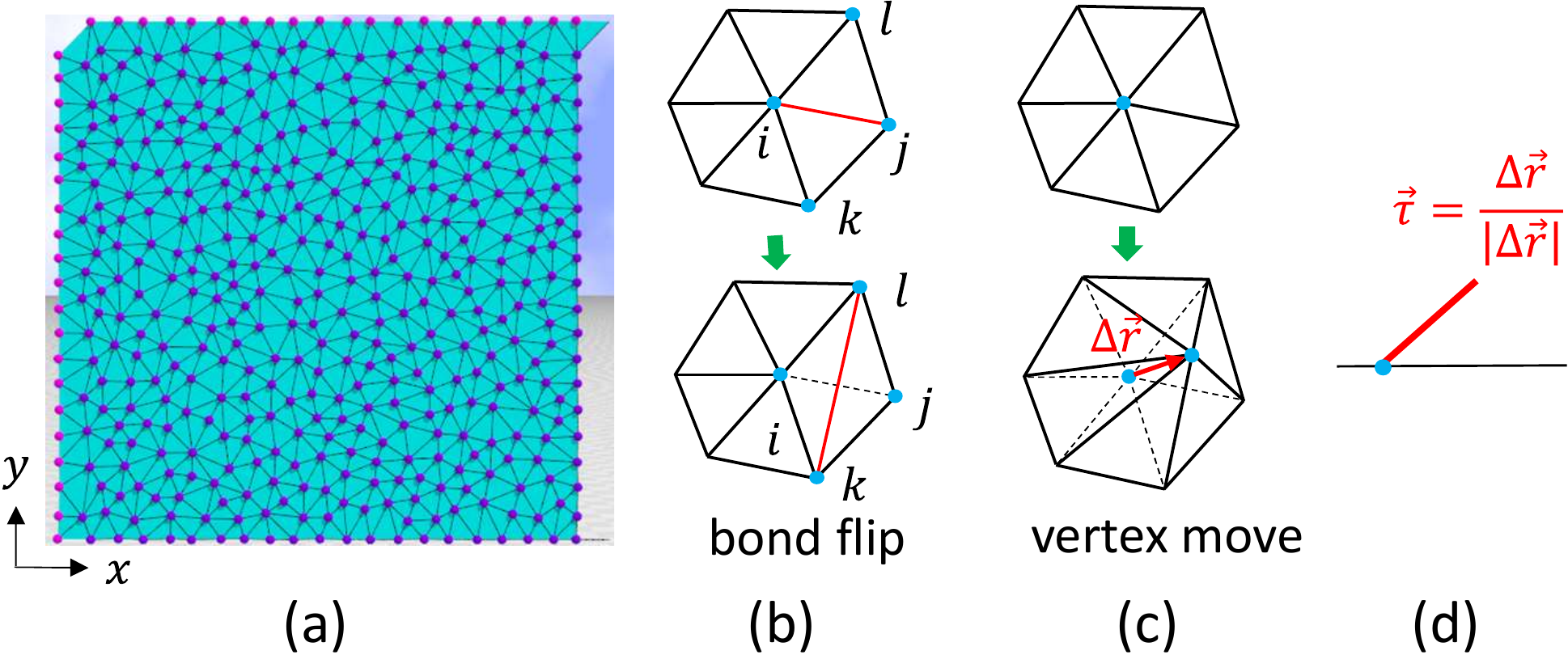}
\caption {(a) Initial configuration of a triangulated lattice with periodic boundary conditions and of size $N\!=\!20^2$, which is the total number of vertices smaller than $N\!=\!50^2$ assumed in the simulations.     (b) An illustration of the bond flip process, which is a basic process of dynamical triangulation $\mathcal{T}$ in $Z_{\rm fluid}$ and allows vertices $i$ and $j$ to diffuse over the surface,  (c) an illustration of a vertex move from $\vec{r}({\rm old})$ to $\vec{r}({\rm new})$, which is a basic process of thermal fluctuations in both $Z_{\rm fixed}$ and $Z_{\rm fluid}$, and (d) definition of IDF $\vec{\tau}$ by using $\Delta \vec{r}=\vec{r}({\rm new})\!-\!\vec{r}({\rm old})$. Vertices on the upper and left boundaries in (a) are identified with those on the opposite boundaries. These vertices fluctuate like the internal vertices in both $Z_{\rm fixed}$ and $Z_{\rm fluid}$, and the boundary bonds also flip like the internal bonds in $Z_{\rm fluid}$.
\label{fig-4} }
\end{figure}
As indicated in $S(\vec{r};\vec{\tau})$, the variable to be updated by MC simulations is the vertex position $\vec{r}$ of the lattice, and IDF $\vec{\tau}$ is defined by the deformation of $\vec{r}$. Therefore, the variables $\{\vec{r}\}$  are the source of  IDF $\{\vec{\tau}\}$, or we simply consider that the vertex position $\vec{r}$ plays a role in the IDF $\vec{\tau}$  for the diffusion anisotropy for $\phi$ and $\psi$.  Here, we should note that vertex move in MC does not correspond to the velocity of the vertex because MC iterations of the variable $\vec{r}$ are not the time evolution of $\vec{r}$. In this sense, vertex fluctuations realized in the MC process do not exactly correspond to thermal  fluctuations of $\vec{r}$ the position of lipids, although the mean value of physical quantity obtained by those fluctuating ensemble configurations of vertices is expected to be a meaningful quantity. Thus, we assume that vertex fluctuations in the MC process represent real thermal fluctuations. Hence, we consider that movement of lipid molecules or lumps of lipids is reflected in the IDF $\vec{\tau}$.

 We use two different types of lattices, fixed connectivity and fluid, of periodic boundary conditions to update the IDFs in MC (Fig.\ref{fig-4}(a)). The partition functions are given by
\begin{eqnarray}
\label{part-funct}
\begin{split}
 &Z_{\rm fixed}=\int \prod_i d\vec{r}_i \exp(-S), \quad (\textrm {fixed model}), \\
 &Z_{\rm fluid}=\sum_{\mathcal T}\int \prod_i d\vec{r}_i \exp(-S), \quad (\textrm {fluid model}).
 \end{split}
\end{eqnarray}
The symbol $\sum_{\mathcal T}$ in $Z_{\rm fluid}$ denotes the sum over possible triangulation ${\mathcal T}$, which is performed by the bond-flip technique \cite{Ho-Baum-EPL1990}  (Fig. \ref{fig-4}(b)). For this bond-flip process of MC updates for triangulation ${\mathcal T}$,  the connectivity of triangles changes, and as a result, vertices freely diffuse over the surface. We should note that the simulation unit is defined by $k_B T\!=\!1$, where $k_B$ and $T$ are the Boltzmann constant and the temperature, respectively. 

 Thus, the numerical procedure is as follows:
\begin{enumerate}
\item[(i)]  Discrete versions of Eq. (\ref{FN-eq-Eucl}) are iterated such that $\phi(t+dt)\leftarrow \phi(t)+dt (\cdots)$ and $\psi(t+dt)\leftarrow \psi(t)+dt (\cdots)$ with the operators $\Laplace_\phi$ and  $\Laplace_\psi$ in Eq. (\ref{FN-eq})
\vspace{-2mm}
\item[(ii)]  In each discrete time step $t \to t+dt$, the variable $\vec{r}$ is updated in MC simulations, as shown in Fig. \ref{fig-4}(c), for sufficiently many sweeps on a  fixed connectivity lattice (fixed model) or dynamically-triangulated fluid lattices (fluid model), which are defined by the partition functions  in Eq. (\ref{part-funct}). At each update of $\vec{r}$, the variable $\vec{\tau}$ is also updated by using $\Delta \vec{r}(=\vec{r}({\rm new})\!-\!\vec{r}({\rm old})$  (Fig. \ref{fig-4}(d)). 
\vspace{-2mm}
\item[(iii)]  Steps (i) and (ii) are repeated until the conditions $\partial \phi/\partial t\!=\! 0$, and  $\partial \psi/\partial t\!=\! 0$ are numerically satisfied in  Eq. (\ref{FN-eq-Eucl}), in which the Laplace operator  $\Laplace=\frac{\partial^2 }{\partial x^2}+\frac{\partial^2}{\partial y^2}$ is replaced by  $\Laplace_\phi$ for $\phi$  and $\Laplace_\psi$ for $\psi$ defined in Eq. (\ref{FN-eq}).
\end{enumerate}

\section{Results}
\begin{figure}[h!]
\centering{}\includegraphics[width=14.5cm]{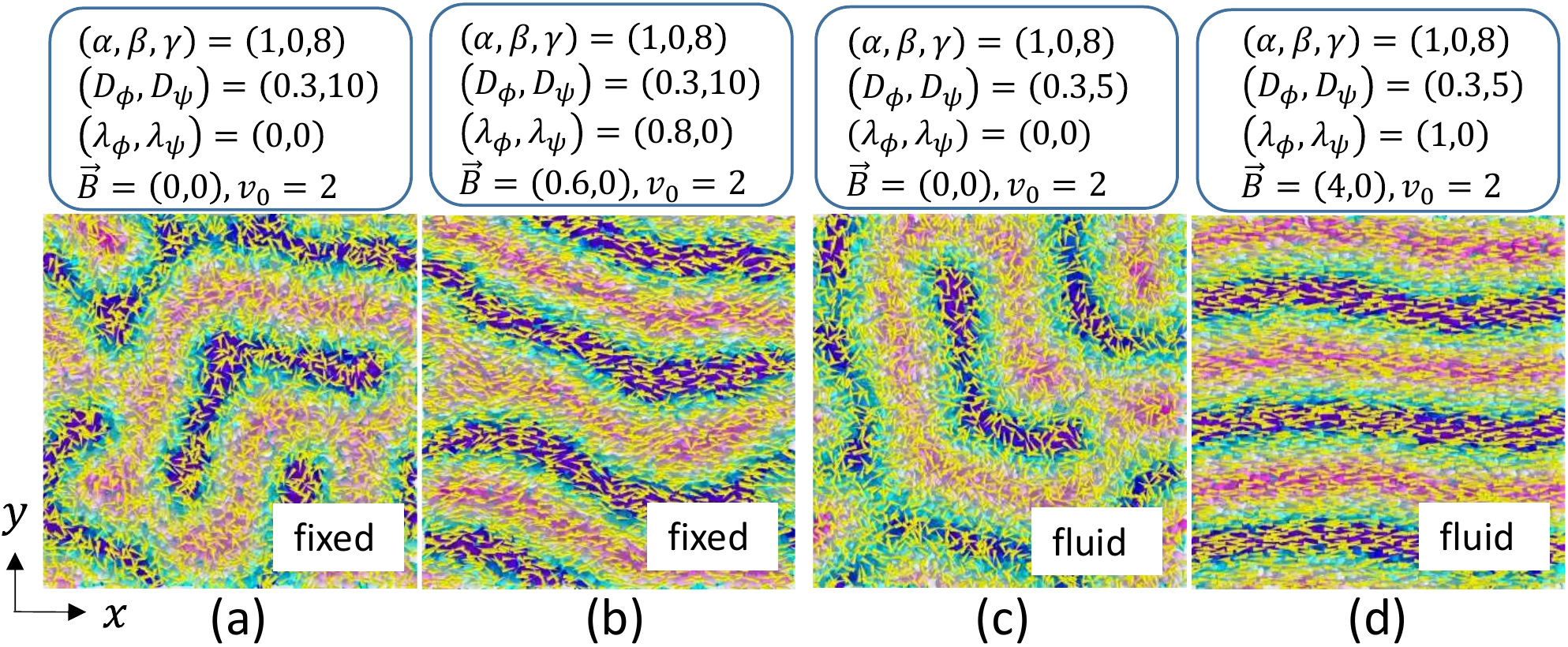}
\caption {(a) Isotropic and (b) anisotropic patterns on a fixed connectivity surface, and (c) isotropic and (d) anisotropic patterns on the fluid surface. The assumed parameters are shown  on the top of the snapshots. The small cones represent IDF $\tau$ indicating the direction of vertex diffusion. The size of the lattice is $N\!=\!50^2$.
\label{fig-5}}
\end{figure}
The simulations are performed on triangulated lattices of periodic boundary conditions of size $N\!=\!50^2$. The results of the fixed connectivity and fluid models are shown in Figs. \ref{fig-5}(a), (b) and  Figs. \ref{fig-5}(c), (d), respectively. The assumed parameters $\alpha,\beta,\gamma$, $D_\phi,D_\psi$ in Eqs. (\ref{FN-eq-Eucl}),   $\lambda_\phi,\lambda_\psi,\vec{B}$ in $S$ of Eq. (\ref{discrete-Hamiltonian}), and 
$v_0$ in $\vec{v}^{\,\phi,\psi}$ in Eq. (\ref{one-dim-Hamiltonian}) are shown in the upper part of each snapshot. We find that isotropic (anisotropic) patterns appear when $\vec{B}\!=\!(B,0)$ with $B\!=\!0$ ($B\!>\!0$) in both fixed and fluid surfaces in Figs. \ref{fig-5}(a), (c)  (\ref{fig-5}(b),(c)), where $\lambda_\phi$ is fixed to be nonzero for anisotropic patterns. The direction of $\vec{\tau}$ controlled by   $\vec{B}$ is closely related to whether the pattern is isotropic or anisotropic. This correspondence between the directions of  $\vec{\tau}$ and the patterns explains the anisotropic Turing patterns. Because anisotropy in $\vec{\tau}$ corresponds to anisotropy in vertex diffusion, and the anisotropic Turing pattern is a consequence of anisotropy in diffusion constants as confirmed in Fig. \ref{fig-2}, we consider that anisotropy in vertex diffusion is reflected in the anisotropy in Turing patterns.

\section{Concluding Remarks}
Preliminary simulation results on anisotropic Turing patterns are presented in this paper. The FitzHugh-Nagumo equation is numerically solved on fixed connectivity and dynamically triangulated surfaces. The vertex position of these lattices plays a role in an internal degree of freedom (IDF) for Finsler geometry (FG) modeling of anisotropy in the Laplacian in the FitzHugh-Nagumo equation, and this IDF is updated in the Metropolis Monte Carlo simulation technique. We find that anisotropy in Laplacian or diffusion anisotropy is connected with anisotropy in vertex diffusion on triangulated lattices via the IDF, implying that Turing patterns can be controlled by the vertex diffusion. Detailed information on the induced anisotropy in diffusion constants of Laplacian should be studied, and further numerical studies are necessary for more detailed information on the relation between Turing patterns and anisotropy in IDF. 

\begin{acknowledgments}
The author H.K. acknowledges Dr. M. Nakayama and Dr. S. Tasaki for comments. The authors acknowledge Dr. Jean-Paul Rieu for suggestions. Numerical simulations were performed on the Supercomputer system "AFI-NITY" at the Advanced Fluid Information Research Center, Institute of Fluid Science, Tohoku University.
\end{acknowledgments}


\nocite{*}
\bibliography{aipsamp}

\end{document}